\documentclass[letterpaper]{article}

\usepackage{amsmath}
\usepackage{graphicx}
\usepackage{geometry}
\usepackage{amssymb}
\usepackage{amsfonts}
\usepackage{eufrak}
\usepackage{yfonts}
\usepackage{amsthm}
\usepackage{geometry}

\begin{document}

\title{ Consensus at Competitive Equilibrium: Dynamic Flow of Autonomous Cars in Traffic Networks}

\author{ Dragoslav D. \v{S}iljak% <-this % stops a space
\thanks{Professor Emeritus of Electrical Engineering, Santa Clara University, Santa Clara, CA 95053, USA. dsiljak@scu.edu}}% <-this % stops a space
%\thanks{Professor Emeritus of Electrical Engineering, Santa Clara University, Santa Clara, CA 95053, USA. dsiljak@scu.edu}}% <-this % stops a space
\maketitle

\begin{abstract}
The objective of this paper is to initiate a qualitative
analysis of dynamic flow in traffic networks by using the competitive
equilibrium model of multiple market systems. A network is modeled as a
dynamic graph where routes (edges) are viewed by drivers (agents) as gross
substitute commodities which they choose by considering the traffic densities
as prices of the individual routes. By borrowing from economic equilibrium
models the notions of gross substitution and homogeneity of excess demand
functions, we will be able to show that the chosen decision rule of the
drivers will lead to a consensus resulting in an even distribution of traffic
density over all routes of the network.
\end{abstract}


\section{Introduction}
The concept of \textit{consensus} was introduced in the 1950s by Arrow, Block and
Hurwicz \cite{Arrow1959}, who showed that under standard conditions for stability of the
competitive equilibrium a nonlinear time-invariant model of multiple-market
system can reach consensus, that is, a unity equilibrium ray where all
normalized prices are equal to each other. At the cost of a more elaborate
analysis, it has been shown \cite{Siljak1976-1} that a time-varying version of the stability
conditions guarantees that consensus will be reached in a time-varying model
of competitive equilibrium. The main objective of this paper is to initiate
a qualitative study of traffic networks using the general competitive analysis
of economic systems \cite{Arrow1959}--\cite{Siljak1978}. The underlying idea is to show that the drivers in a traffic network (like economic agents in a multiple market
system), when choosing alternative routes based solely on the density of
traffic in the corresponding routes, can reach an even distribution
(consensus) of traffic density over the network. We represent traffic
networks by dynamic graphs which were introduced in \cite{Siljak2008} to consider graphs as
dynamic systems. By imitating competitive market models, we then consider
edges of a dynamic graph as commodities (goods) and traffic densities on the
edges as their prices. 

The two basic assumptions in dynamic graph models of traffic networks are
borrowed from the competitive economic analysis. \ First, we consider the
edges of a graph as \textit{gross substitute} \textit{commodities, }meaning
that if two edges are substitutes (for each other) then an increase in the
density of one edge results in an increase in the demand for the other edge.
Second, we assume that the choice of edges (routes) by the drivers depends on
the normalized densities of the routes, and not on the absolute
densities of the individual routes. This fact implies that \textit{the
excess demand functions are homogeneous of degree zero}. With a few
additional technical assumptions we will be able to show that the agents will
bring about a desired state of equilibrium of the traffic flow.

There is a large body of work on flows in networks and, in particular, in
traffic networks. Informative recent surveys and discussions of models and
control design methods for traffic flows are provided in the papers \cite{Darbha1999}--\cite{Tyagi2008}.
 In general, the models of traffic networks treat a stream of vehicles very
much like water in a water distribution network, or gas in a gas distribution
network, etc. Individual drivers of the vehicles do not have decision power,
but follow the rules set by the parameters of the model reflecting the physics
of the flow, or are directed by external feedback control. As distinct from
this approach, we consider traffic networks as interconnected systems with
implicit assumption of dynamic coupling among traffic densities on all the
routes. In our model, the drivers behave as economic agents who choose in a decentralized way individual routes solely on the basis of density of the traffic in any given
route. The densities define a state of the network which is represented by a
nonlinear time-varying \textit{dynamic system}. In the course of our analysis we will establish existence of system motions and attraction of an
unit equilibrium ray in the state space of the network. By converging to the
equilibrium ray the agents reach the \textit{consensus} resulting in an even
distribution of traffic over the network.

\section{Competitive Dynamic Graphs}
We consider a \textit{weighted directed graph} (or simply, \textit{graph})
$D=($V$,$E$)$ which is an ordered pair, where V = $\{$v$_{1},$v$_{2},\ldots
,$v$_{n}\}$ is a nonempty and finite set of \textit{vertices}
(\textit{points)} and E = $\{e_{1},e_{2},\ldots,e_{m}\}$ is a family of
elements of the Cartesian product V $\times$ V, called \textit{edges}
(\textit{lines}). Each edge $($v$_{j},$v$_{i})$ is oriented from v$_{j}$ to
v$_{i}$, and is assigned a real number $e_{k}$, the \textit{weight} of the
edge. \ Graph $D$ is a \textit{multigraph }if there are two or more
\textit{parallel} \textit{edges} that join the same pair of vertices.
\ Parallel edges are considered distinct and are labeled individually. A
graph without parallel edges is a \textit{simple graph}, in which case E is a set.

Dynamic graphs were defined in \cite{Siljak2008} by setting up a linear graph space
$\mathfrak{D}$ and describing motions of graphs by a one-parameter group
$D(t;t_{0},D_{0})$ of transformations of the space $\mathfrak{D}$ into itself.
 Since dynamic graphs so defined are isomorphic to dynamic \textit{adjacency
(interconnection) matrices}, an $m$-dimensional linear space $\mathfrak{E}$ of
adjacency matrices was defined, where the analysis of dynamic graphs has been
carried out as motions $E(t;t_{0},E_{0})$ of adjacency matrices $E=(e_{ij})$
in terms of edge weight vectors $e\in\mathfrak{E}$ defined over the field
$\mathfrak{F}$ of real numbers. By choosing $\mathbb{R}^{m}$ for
$\mathfrak{E}$, a dynamic adjacency matrix $\mathbf{E}$ was described by
differential equation%
\begin{equation}
\mathbf{E:\,\,\ }\dot{e}=g(t,e),
\end{equation}
where function $g:\mathcal{T\,}\mathbb{\times\,R}^{m}\rightarrow\mathbb{R}%
^{m}$ was assumed to be sufficiently smooth, so that solutions $e(t;t_{0}%
,e_{0})$ of $\mathbf{E}$\ exist and are unique for all $(t_{0},e_{0}%
)\in\mathcal{T}\times\mathbb{R}^{m}$and $t\in\mathcal{T}\,_{0}$, where
$\mathcal{T\,}$is the time interval $(\tau,+\infty)$, $\tau$ is a number or
symbol $-\infty$, and $\mathcal{T}_{0}$ is the semi-infinite time interval
$[t_{0},+\infty).$ \ How the model $\mathbf{E}$ of minimal dimension $m$ can
be obtained from a given graph $D$ via the fundamental interconnection matrix
$\bar{E}=(\bar{e}_{ij})$ has been explained in \cite{Siljak2008}. 

To introduce a competitive model of multi-agent traffic networks, we use the
theory of \textit{competitive equilibrium} of multiple market systems (e.g. 
\cite{Arrow1959}, \cite{Arrow1971}, and \cite{Siljak1978}). We recall that economic agents meet at the market place to exchange
goods making their independent decisions solely on the basis of the prices
which they cannot control and take them as given. This is a characteristic
of a perfectly competitive environment, where actions of economic agents can
be all mutually compatible and carried out simultaneously. In a graph
representing a traffic network, we consider edges $($v$_{j},$v$_{i})$ as goods
and weights $e_{ij}$, which represent the densities of traffic flows through
the corresponding edges, as prices of these goods. Then, the agents
utilizing the network select edges on the basis of their weights and under
certain conditions can reach a traffic equilibrium.

To define a competitive graph $\mathbf{D,}$ we recall the \textit{gross
substitute case} of multiple market systems \cite{Siljak2008} in which all goods are
substitutes. We further recall that the gross substitute case is defined in
terms of the excess demand function $g(t,e)$ as follows:

$\mathbf{Assumption}$ $(\mathbf{A}_{1}).$\textbf{ \ }Function $g(t,e)$ belongs
to the class of gross substitute functions%
\begin{equation}%
\begin{tabular}
[c]{l}%
$\mathcal{\bar{G}}\text{:}\mathcal{\ }\text{ }g_{i}(t,e^{\prime})\leq
g_{i}(t,e^{\prime\prime}),\,\ \ \forall(t,e^{\prime}),(t,e^{\prime\prime}%
)\in\mathcal{T\ }\mathbb{\times\,R}_{+}^{m},\,\ \text{\ }\forall
i\in\mathbb{M}$\\
$\ \ \ \ \ \ \ e_{i}^{\prime}=e_{i}^{\prime\prime},$ $\ \ e_{j}^{\prime}\leq
e_{j}^{\prime\prime},\,\ \ \forall j\in\mathbb{M},i\neq j$,
\end{tabular}
\end{equation}
where $\mathbb{R}_{+}^{m}=\{e\in\mathbb{R}^{m}:e_{i}\geq0,i\in\mathbb{M\}}$
denotes the nonnegative orthant in $\mathbb{R}^{m},$ and $\mathbb{M}%
=\{1,2,\ldots,n\}.$

Since we are interested in compatibility of the decisions of economic agents
we are necessarily concerned with the difference of demand and supply of the
individual goods on the market. The value of the function $g(t,e)$ tells us
what the excess demand would be if all economic agents carried out\ their
prefered actions at price $e.$ Furthermore, if price of one good goes up
while all other prices stay constant, there will be excess demand for all
goods whose prices has remained constant.

To shed more light on the modeling problem we recall \cite{Siljak2008} the notion of an
\textit{equilibrium graph} $D^{e}$ defined by $D(t;t_{0},D^{e})=D^{e}$.  A
constant weight vector $e^{e}$, which corresponds to $D^{e}$, defines the
\textit{equilibrium adjacency matrix} $E^{e}$ by the equivalent condition
\begin{equation}
e(t;t_{0},e^{e})=e^{e},~\ \ \ \forall t\in\mathcal{T\,}.
\end{equation}
In the context of multiple market systems, $e^{e}$ is a price vector at which
all agents achieve what they want, and there is no action to cause a change in
$e$.  At $e^{e}$ decisions of all agents are compatible and they can be
carried out simultaneously.  A change in price vector $e$ from the
equilibrium price $e^{e}$ indicates$\ $incompatibility in the decisions of the
agents; this is the familiar notion of the "law of supply and demand."

To illustrate our modeling proposition, let us consider the \textit{linear
constant model} of multiple market systems \cite{Arrow1966}. Then, a competitive dynamic
graph $\mathbf{D}$ is described via adjacency matrix as
\begin{equation}
\mathbf{E}_{L}:\dot{e}=Ae+b
\end{equation}
where $e=(e_{1},e_{2},\ldots,e_{m})^{T}\in\mathbb{R}_{+}^{m}$ is the edge
vector standing for the price vector, $A=(a_{ij})$ is a constant $m\times m$
matrix with sign-pattern%
\begin{equation}%
\begin{array}
[c]{ll}%
a_{ij} & <0,\,\ i=j\\
& \geq0,\,\ i\neq j
\end{array}
\end{equation}
and $b=(b_{1},b_{2},\ldots,b_{m})^{T}\in\mathbb{R}_{+}^{m}$ is a nonnegative
constant vector $(b_{i}\geq0,i\in\mathbb{M)}$. \ The fact that the
off-diagonal elements of matrix $A$ are nonnegative makes the matrix $A$ a
\textit{Metzler matrix} \cite{Arrow1983}, implying further that $Ae+b\in
\mathcal{\bar{G}}$ defined in assumption $(A_{1}).$ \ In economic terms, we
have a gross substitute case where all edges are substitutes; the dynamic
system $\mathbf{E}_{L}$ is a linear constant model of competitive equilibrium
$e^{e}=-A^{-1}b$ [4].

Let us now illustrate the idea of a competitive dynamic graph in modeling of
traffic networks by a simple example involving a dynamic multigraph with two
vertices and two parallel edges as shown in Fig. 1.
I[h]
\begin{figure}
\begin{center}
\includegraphics[width=2.5in]{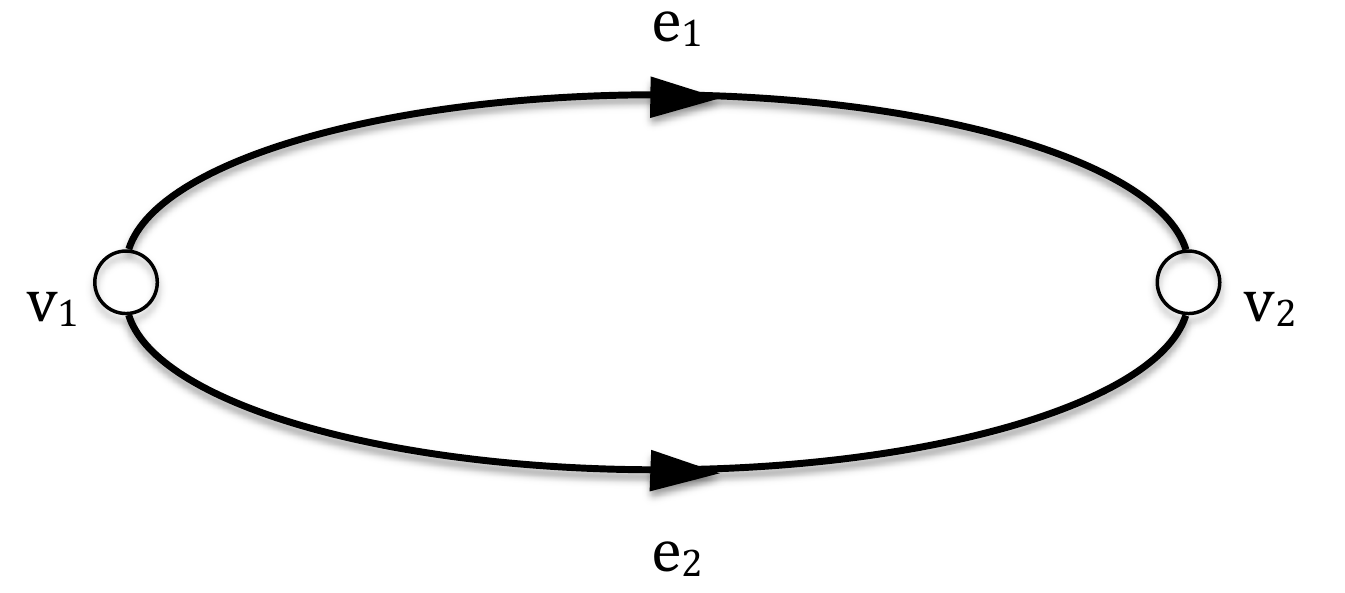}
\caption{ A multigraph with two parallel routes.}
\end{center}
\end{figure}
The graph represents a traffic network where drivers can get
from vertex v$_{1}$ to vertex v$_{2}$ by either of the two edges, where the
weights $e_{1}$ and $e_{2}$ represent the density of traffic in the respective
edges. \ The instantaneous values of the density of flows $e_{1}(t)$ and
$e_{2}(t)$ are determined by the differential equations%
\begin{equation}%
\begin{array}
[c]{ll}%
\mathbf{E}_{L}\mathbf{:} & \dot{e}_{1}=a_{11}e_{1}+a_{12}e_{2}+b_{1}\\
& \dot{e}_{2}=a_{21}e_{1}+a_{22}e_{2}+b_{2}%
\end{array}
\end{equation}
Now, if the density $e_{1}(t)$ rises above the equilibrium value $e_{1}^{e}$,
the drivers would switch to road $e_{2}$, which is a substitute to $e_{1}$.
\ The switch would decrease the density $e_{1}(t)$ while, at the same time,
cause the density $e_{2}(t)$ to rise due to positivity of the coefficient $a_{12}$ (the matrix $A$ of
$\mathbf{E}_{L}$ is a Metzler matrix). \ Then, the drivers entering the vertex
v$_{1}$ would begin to prefer the road $e_{1}$ over $e_{2}$, and so on. \ The
fact that this type of price adjustment process converges to the competitive
equilibrium of a multiple market system with unfailing regularity is the most
visible part of the Adam Smith ``invisible hand." What we want to do in this paper is to capitalize
on this remarkable feature of competitive economic systems and propose a
similar model for traffic flow networks.

The adjustment process, which was described in the example, goes on throughout the traffic network. The drivers constitute completely decentralized controllers (\cite{Siljak1978}, \cite{Siljak1991}), which are continuously choosing the less traveled roads resulting in a balanced traffic density in all routes of the network; the drivers collectively constitute a ``swarm" over the network.  For the adjustment process to succeed, it is essential that the drivers obtain the information about density distribution over the routes of the network. At present, this information is provided either by GPS Maps situated on the dashboard of the cars, or by smart phones.

The class $\mathcal{\bar{G}}$ of functions $g(t,e)$ was introduced by M\"{u}ler \cite{Muler1926} to serve as a basis for the \textit{comparison} \textit{principle} in
the theory of differential inequalities (Kamke \cite{Kamke1932}; Wa\v{z}evski \cite{Wa1950};
Lakshmikantham and Leela \cite{Lakshmikantham1969}; and Ladde \cite{Ladde1976}). The fact
that the same class $\mathcal{\bar{G}}$\ represents excess demand functions of
gross\textit{ }substitute goods in the \textit{competitive equilibrium models}
of market systems in mathematical econoimics (e. g., Arrow and Hahn \cite{Arrow1971}), was
first recognized in (\v{S}iljak \cite{Siljak1976-2}). This recognition has been exploited
in obtaining new results in stability analysis of models in as diverse fields
as population biology (\v{S}iljak \cite{Siljak1978}, \cite{Siljak1975}; Ladde \cite{Ladde1976};
Ikeda and \v{S}iljak \cite{Ikeda1980}), arms race (\v{S}iljak \cite{Siljak1976-2}, \cite{Siljak1977}), chemical
systems (Ladde \cite{Ladde1976}, Ladde \cite{Ladde1976-2}, Maeda et al. \cite{Maeda1978}), compartmental models (Ladde \cite{Ladde1976}; Jacquez \cite{Jacquez1993}), and
large-scale systems (\v{S}iljak \cite{Siljak1978}; Michel and Miller \cite{Michel1977}; Lakshmikantham
et al. \cite{ Lakshmikantham1991}; \v{S}iljak \cite{Siljak1991}; Martynyuk \cite{ Martynyuk2002}, \cite{Martynyuk2007}).

\section{Positivity}
For either physical or conceptual reasons, variables of dynamic systems are
required to be nonnegative or strictly positive. \textit{Positive dynamic
systems} have been introduced in economics as mathematical models of multiple
market systems where prices start and stay positive
evolving within the positive orthant for all time.  Similarly, in traffic
networks, the densities of traffic on routes of the network are necessarily
nonnegative variables. Within the context of dynamic graphs (\v{S}iljak \cite{Siljak2008}), this restriction means that if a graph (network) trajectory starts with
nonnegative weights representing the traffic densities, it stays with
nonnegative weights for all future time. A positive dynamic graph
$\mathbf{D}$ and the corresponding dynamic matrix $\mathbf{E}$ belong to a
broad and extensively studied class of \textit{positive dynamic systems}
(e.g., Arrow and Hahn \cite{Arrow1971}; \v{S}iljak \cite{Siljak1978}, \cite{Siljak2008}, \cite{Siljak1976-2};
Ladde \cite{Ladde1976}; Krasnosel'skii \cite{Krasnoselskii1966}). Recently, breaking with tradition, these type of systems were
termed \textit{nonnegative systems} (Haddad et al. \cite{Haddad2004}, \cite{Haddad2005}). We note
that variables of competitive systems are nonnegative, and competitive systems
are positive dynamic systems (see Remark 1 below), but they are only a (large)
subclass of such systems. To show this fact, we start with

\textbf{Definition 1.} \ Adjacency matrix $\mathbf{E}$ is a positive dynamic
system if $e_{0}\in\mathbb{R}_{+}^{m}$ implies $e(t;t_{0},e_{0})\subset
\mathbb{R}_{+}^{m}$ for all $t\in\mathcal{T}_{0}$, that is, $\mathbb{R}%
_{+}^{m}$ is an \textit{invariant set} of $\mathbf{E}$. \ 
To establish positivity of dynamic graphs we need the following:

$\mathbf{Assumption}$ $(\mathbf{A}_{2}).$\textbf{ \ }Function $g(t,e)$ belongs
to the class
\begin{equation}%
\begin{tabular}
[c]{l}%
$\mathcal{N}\text{: \ }g_{i}(t,e_{1},\ldots,e_{i-1},0,e_{i+1},\ldots
,e_{m})\geq0,\,\ \ \forall(t,e)\in\mathbb{R\times R}_{+}^{m},\,\ \text{\ }%
\forall i\in\mathbb{M}$\\
$\ \ \ \ \ e_{j}\geq0,\,\ \forall j=1,\ldots,i-1,i+1,\ldots,m.$%
\end{tabular}
\end{equation}
Now, we prove the following:

\textbf{Theorem 1.} \ Under assumption $(A_{2})$ the adjacency matrix
$\mathbf{E}$\textbf{ }is a positive dynamic system.

\textbf{Proof.} \ We follow \cite{Krasnoselskii1966} and associate with dynamic
matrix $\mathbf{E}$ the auxiliary system%
\begin{equation}%
\begin{array}
[c]{ll}%
\mathbf{E}_{\varepsilon}: & \dot{e}_{1}\,\ =g_{1}(t,e_{1},\ldots
,e_{m})+\varepsilon\\
& \dot{e}_{2}\,\ =g_{2}(t,e_{1},\ldots,e_{m})+\varepsilon\\
& \cdots\cdots\cdots\cdots\cdots\cdots\cdots\cdots\cdots\\
& \dot{e}_{m}=g_{m}\left(  t,e_{1},\ldots,e_{m}\right)  +\varepsilon
\end{array}
\end{equation}
where $\varepsilon>0$. \ Set $\mathbb{R}_{+}^{m}$ is invariant with respect to
the auxiliary system $\mathbf{E}_{\varepsilon}$ since at every point of
intersection of a solution of $\mathbf{E}_{\varepsilon}$ with the boundary of
$\mathbb{R}_{+}^{m}$, the components of the solution (which are zero) are
increasing. \ When we let $\varepsilon\rightarrow0$, the solutions of the
auxiliary system $\mathbf{E}_{\varepsilon}$ become solutions of the original
system $\mathbf{E}$. \ Since $\mathbb{R}_{+}^{m}$ is a closed set, the theorem
follows. 

\textbf{Remark 1.} \ We note that $(A_{2})$ implies $(A_{1})$ provided
$g(t,0)=0$ for all $t\in\mathcal{T}$ (Ladde, 1976), which\ further implies
that gross substitute assumption ($A_{1}$), with the origin as an equilibrium
point of $\mathbf{E}$\textbf{, }implies that dynamic adjacency matrix
$\mathbf{E}$ is a positive dynamic system. \ This fact, in turn, implies that
competitive dynamic systems discussed in the last paragraph of Section 2, are
positive systems, while the opposite implication is not true in general.

\section{Existence and Attraction}
As it is standard in competitive equilibrium analysis \cite{Arrow1971}, we assume that
the economic agents choose goods depending on their relative prices and not on
the prices given in terms of a medium of exchange (e.g., money). \ If we have
a multiple market system described by the equation%
\begin{equation}
\mathbf{\bar{E}:\,\mathbf{\,}}\mathtt{\dot{e}}~=\mathtt{g}(t,\mathtt{e}),
\end{equation}
where $\mathtt{e}$ $=(\mathtt{e}_{0},\mathtt{e}_{1},...,\mathtt{e}_{m})$ is
the price vector of $m+1$ goods, and $\mathtt{g}:\mathcal{T\,}\mathbb{\times
\,R}^{m+1}\rightarrow\mathbb{R}^{m+1}$ is the excess demand function, we can
arbitrarily set one of the prices to 1, say $\mathtt{e}_{0}=1,$ and treat
other prices as \textquotedblright normalized prices.\textquotedblright\ \ A
good having the price $\mathtt{e}_{0}$ need not be money, but instead any good
that serves as \textit{num\'{e}raire, }a measuring stick, with prices of all
other goods being expressed in terms of its price. Since $\mathtt{e}_{0}=1$,
we write $e_{i}=\mathtt{e}_{i}/\mathtt{e}_{0}$ and from the system
$\mathbf{\bar{E}}$ expressed in terms of \textit{non-normalized} prices
$\mathtt{e}_{0},\mathtt{e}_{1},...,\mathtt{e}_{m}$ we obtain the original
system $\mathbf{E}$ in terms of the \textit{normalized }prices $e_{1}%
,e_{2},...,e_{m}.$ Then, the components of excess demand functions are defined
as
\begin{equation}
\mathtt{g}_{i}(t,\mathtt{e}_{0},\mathtt{e}_{1},...,\mathtt{e}_{m}%
)=\mathtt{g}_{i}(t,1,e_{1},e_{2},...,e_{m})
\end{equation}
implying that the excess demand functions in terms of normalized prices are
given as%
\begin{equation}
g_{i}(t,e_{1},e_{2},...,e_{m})=\mathtt{g}_{i}(t,1,e_{1},e_{2},...,e_{m})
\end{equation}
In the context of traffic networks, we take the density in one of the routes
as a num\'{e}raire, which will serve as a reference for all other route
densities taken as prices of the edges (goods). \ We hasten to add that the
density of the num\'{e}raire need not be constant, but may be varying in time.
We need the following \cite{Siljak1976-1}:

$\mathbf{Assumption}$ $\mathbf{A}_{3}.$\textbf{ \ }Function $g(t,e)$ belongs
to the class of positive homogeneous functions of degree zero
\begin{equation}
\mathcal{H}\text{: \ }g(t,\lambda e)=g(t,e),~\ \ \ \ \forall\lambda>0.
\end{equation}

To consider existence of solutions $e(t;t_{0},e_{0})$ of  $\mathbf{E}$\textbf{
}we need a stronger version of assumption $(A_{1})$:

$\mathbf{Assumption}$ $\mathbf{(A}_{1}^{\prime}).$\textbf{ \ }Function
$g(t,e)$ belongs to the class of strong gross substitute functions%
\begin{equation}%
\begin{tabular}
[c]{l}%
$\mathcal{G}\text{:}\mathcal{\ }\text{ }g_{i}(t,e^{\prime})<g_{i}%
(t,e^{\prime\prime}),\,\ \ \forall(t,e^{\prime}),(t,e^{\prime\prime}%
)\in\mathcal{T\,\ }\mathbb{\times\,}\mathcal{C},\,\ \text{\ }\forall
i\in\mathbb{M}$\\
$\ \ \ \ \ \ \ e_{i}^{\prime}=e_{i}^{\prime\prime},$ $\ \ e_{j}^{\prime}%
<e_{j}^{\prime\prime},\,\ \ i\in\mathbb{M}$,$~\ j\in\mathbb{M-J}$,
\end{tabular}
\end{equation}
where
\begin{equation}
\mathcal{C}=\{e\in\mathbb{R}_{+}^{m}:e_{i}>0,\,\ \forall i\in\mathbb{M\}}%
\end{equation}
is an open cone in $\mathbb{R}_{+}^{n}$, and $\mathbb{J}$ is a nonvoid subset
of $\mathbb{M}$.

We also need the following :

$\mathbf{Assumption}$ $(\mathbf{A}_{4}).$\textbf{ \ }There exists an
equilibrium $e^{e}\in\mathcal{C}$ as a solution of the equation $g(t,e)=0$.

In terms of normalized densities, we obtain the following theorem concerning
the equilibrium:

\textbf{Theorem 2. \ }Under the assumptions:%
\begin{equation}%
\begin{array}
[t]{ll}%
\left(  A_{1}^{\prime}\right)  & g(t,e)\in\mathcal{G}\\
\left(  A_{3}\right)  & g(t,\lambda e)=g(t,e),\forall\lambda>0\\
\left(  A_{4}\right)  & \exists e^{e}\in\mathcal{C}\text{:~\ }g(t,e^{e}%
)=0,\forall t\in\mathcal{T}%
\end{array}
\end{equation}
there exists a unique equilibrium ray
\begin{equation}
\mathcal{E}=\{e^{e}\in\mathcal{C}:e^{e}=\lambda\mathbf{e}\}
\end{equation}
$\mathtt{of}$ \textbf{E}, where $\mathbf{e}$ $=\{1,1,...,1\}$\ $\in
\mathbb{R}_{+}^{m}$\ and $\lambda$ is a positive number.

\textbf{Remark 2. }Theorem 2 is interesting because it says that given
sufficiently long time, the densities on all routes of the network will all be
equal to each other. This result was first obtained for a nonlinear
time-varying version (\v{S}iljak \cite{Siljak1976-1}) of the Richardson's model of the
arms race. This phenomenon was termed \textquotedblright
consensus\textquotedblright\ in recent studies of multi-agent systems (e.g.,
Ren and Beard \cite{Ren2008}), where the differences (system $\mathbf{L}$ in Section 5
below) instead of ratios (system $\mathbf{E}$) of state variables were considered.

Using Theorem 2, we establish the existence result regarding the motion of
densities over time (\v{S}iljak \cite{Siljak1976-1}, Ladde and \v{S}iljak \cite{Ladde1976}):

\textbf{Theorem 3. \ }If function $g(t,e)$ satisfies the assumptions $\left(
A_{1}^{\prime}\right)  ,$ $\left(  A_{3}\right)  $ and $\left(  A_{4}\right)
$ of Theorem 2, then there exists a solution $e(t;t_{0},e_{0})$ of \textbf{E
}for any $(t_{0},e_{0})\in\mathcal{T\,\times C}$ and all $t\in\mathcal{T}%
_{0}.$

Both Theorems 2 and 3 are proved in the Appendix under weaker conditions.
\ From the proof of Theorem 2, it is clear that the assumptions further imply
that all solutions are bounded on $\mathcal{T}_{0}$ for all $(t_{0},e_{0}%
)\in\mathcal{T\,}\times\mathcal{C}$. \ Moreover, the solutions are positive
and the open cone $\mathcal{C}$ is an invariant set. \ That is, the solutions
have the following property:

\textbf{Property }$(\mathbf{P}_{1})$\textbf{. \ }$(t_{0},e_{0})\in
\mathcal{T\,\times C~\ }\Rightarrow~e(t;t_{0},e_{0})\subset\mathcal{C}$,
$\forall t\in\mathcal{T}_{0}.$

The same set of assumptions of Theorem 2 imply, not only that the cone
$\mathcal{C}$ is invariant, but that it is also a region of attraction of the
equilibrium ray $\mathcal{E}$. \ Formally, we define the following

\textbf{Property }$(\mathbf{P}_{2})$\textbf{. \ }$(t_{0},e_{0})\in
\mathcal{T\,\times C~\ }\Rightarrow~\lim_{{\large t\rightarrow+\infty}%
}d[e(t;t_{0},e_{0}),\mathcal{E}]=0,$ where%
\begin{equation}
d(e,\mathcal{E})=\inf_{{\Large e}^{{\Large e}}{\Large \in}\mathcal{E}%
}\{\parallel e-e^{e}\parallel_{M}\},
\end{equation}
and $\parallel e\parallel_{M}=\sup_{{\Large i\in}\mathbb{M}}\{\parallel
e_{i}\parallel\}.$

In the Appendix, we prove a slightly stronger result than the following:

\textbf{Theorem 4. \ }If the function $g(t,e)$ satisfies the assumptions
$\left(  A_{1}^{\prime}\right)  ,$ $\left(  A_{3}\right)  $ and $\left(
A_{4}\right)  $ of Theorem 2, then the solutions $e(t;t_{0},e_{0})$ of
\textbf{E }have the property $(P_{2})$.

\section{Vertex Dynamics}
So far, we have considered graphs with dynamic edges, leaving the edges static
(constant in time). The reversed situations, where the dynamics is
restricted to the vertices of a graph while the edges were state and/or
time-dependent, have been extensively studied in the context of
\textit{connective stability} of interconnected systems (\v{S}iljak \cite{Siljak1978},
\cite{Siljak1975}; Lakshmikantham et al. \cite{Lakshmikantham1991}; Martynyuk \cite{Martynyuk2002}, \cite{Martynyuk2007}). A graph with
vertex dynamics can be considered as a vertex system
\begin{equation}
\mathbf{V:~}\dot{v}=f(t,v)
\end{equation}
where $v\in\mathbb{R}^{N}$ is the state of \textbf{V }and function
$f:\mathcal{T\,}\mathbb{\times\,R}^{n}\rightarrow\mathbb{R}^{n}$ can be
required to satisfy the same conditions as function $g(t,e)$ of \textbf{E}.
The system \textbf{V} can be used to represent a nonlinear time varying
interaction in a team of $n$ agents (e.g., unmanned vehicles in the air,
water, and on the ground) that share information to reach consensus via often
unreliable communication channels with uncertain topology (see the papers by
Lin et al. \cite{Lin2004}, \cite{Lin2007}; Fax and Murray \cite{Fax2004}; Olfati-Saber and Murray \cite{Saber2004}; Li and Jiang\cite{Li2008}; and a recent book by Ren and Beard \cite{Ren2008}). The most popular consensus algorithm is
\begin{equation}
\mathbf{L:~}\dot{v}=-\sum_{j\neq i}^{n}e_{ij}[v_{i}(t)-v_{j}(t)],~\ \ \ i\in
\mathbb{N},
\end{equation}
where $e_{ij}$'s are the constant elements of the $n\times n$ adjacency matrix
$E=(e_{ij}),$ and $\mathbb{N}=\{1,2,\ldots,n\}.$ \ The purpose of the
algorithm is to achieve the consensus by driving the information state of each
individual agents toward the state of its neighbors, that is, to have
$\left\Vert v_{i}-v_{j}\right\Vert \rightarrow 0,$ as $t\rightarrow\infty,$ for
all $i\in\mathbb{N}.$ \ \
Let us propose that the consensus algorithm be described by a nonlinear time
varying system $\mathbf{V.}$ \ Then, the desired consensus result is obtained
from Theorem 4 as

\textbf{Theorem 5. \ }If the function $f(t,v)$ satisfies the assumptions
$\left(  A_{1}^{\prime}\right)  ,$ $\left(  A_{3}\right)  $ and $\left(
A_{4}\right)  $, then the agents reach consensus defined by the property
\begin{equation}
(P_{2}):(t_{0},v_{0})\in\mathcal{T\,\times C~\ }\Rightarrow~\lim
_{{\large t\rightarrow+\infty}}d[v(t;t_{0},e_{0}),\mathcal{E}]=0,
\end{equation}
where $\mathcal{E}=\{v^{e}\in\mathcal{C}:v^{e}=\lambda\mathbf{e}\}$ and
$v^{e}$ is the equilibrium state of $\mathbf{V}$.

The essential difference between the consensus defined for the system
$\mathbf{L}$ and our consensus guaranteed by Theorem 5 is that we derived the
consensus via scaling of the information state by a num\'{e}raire, which, for
example, can be taken as the information state of the team leader. \ A
possible advantage of the scaled consensus over the difference one should be
explored in future research by a more refined analysis followed by simulation
and practical implementation.

Communication exchange among the individual pairs of agents may be disrupted
due to unreliable communication topology (Ren and Beard \cite{Ren2008}) resulting in a
loss of consistency, accuracy, and completeness of of information necessary to
achieve consensus.  A natural way to capture these type of structural
perturbations is to allow the adjacency matrix $E$ to be time-varying and
state dependent ($E:\mathcal{T\,}\mathbb{\times\,R}^{n}\rightarrow
\mathbb{R}^{n\times n})$ within the concept of \textit{connective stability
}(\v{S}iljak \cite{Siljak1978}) and derive conditions under which we are guaranteed
consensus despite arbitrary disconnections and connections of communication
links between the agents. In the case of algorithm (system ) $\mathbf{L}$ we
need only replace each constant elements $e_{ij}$ by a function $e_{ij}%
:\mathcal{T\,}\mathbb{\times\,R}^{n}\rightarrow\mathbb{R}_{+}.$ When we deal
with the nonlinear time-varying model $\mathbf{V}$, we need to specify the
structure of components $f_{i}$ of the function $f(t,v)$ as%
\begin{equation}
f_{i}(t,v,e)=f_{i}(t,v_{i},e_{i1}v_{1},e_{i2}v_{2},\ldots,e_{in}%
v_{n}),~\ \ \ i\in\mathbb{N}\text{.}%
\end{equation}
There is a large number of results which exploit this structural form within
the concept of connective stability involving stochastic elements, time-delays,
discontinuous nonlinearities, and parametric uncertainties (\v{S}iljak \cite{Siljak1978};
Lakshmikantham et al. \cite{Lakshmikantham1991}; Malikov and Matrosov \cite{Malikov1998}; Stipanovi\'{c} and
\v{S}iljak \cite{Stipanovic2000}, \cite{Stipanovic2004}; Ladde \cite{Ladde2002}; Chandra and Ladde \cite{Chandra2004}). The results can be carried over to multi-agent systems in a natural
way (\v{S}iljak \cite{Siljak2008}).

Recently, dynamic graphs have been introduced which involve both the vertex
and edge dynamics (\v{S}iljak \cite{Siljak2008}) in an composite system configuration%
\begin{equation}%
\begin{tabular}
[c]{ll}%
\textbf{V}$\mathbf{:}$ & $\dot{v}=f(t,v,e)$\\
\textbf{E}$\mathbf{:}$ & $\dot{e}=g(t,e,v)\text{.}$%
\end{tabular}
\ ,
\end{equation}
which was studied within the new concept of \textit{dynamic connective
stability}. \ A\ multi-agent system has been considered as an add-on adaptive
control device based on the dynamic adjacency matrix \textbf{E} to drive the
interconnections of a composite system \textbf{V\&E }to\textbf{ }a preassigned
state which is required to be its stable equilibrium. \ This opened up a new
approach to control of complex systems, which elevated the role of
interconnections (edges) to the same level as subsystems (vertices) in shaping
the performance of interconnected systems. \ In future research, we plan to
explore how the unified vertex and edge dynamics can be used to describe and
analyze more refined models of traffic networks.

\section{Conclusion}
Our main objective in this paper was to model a traffic network as a multiple market system in the context of competitive equilibrium analysis.  Individual routes were considered as commodities and their traffic densities as prices.  Now, drivers become economic agents who choose alternative routes based solely on their prices and can reach a traffic consensus (unity vector) where normalized densities obtain the same value.  Traffic network was modeled as a dynamic graph and Lyapunov's theory was used to establish stability of its consensus.  In the proposed setting, a great deal can be done to improve the competitive model of traffic networks and come up with new and interesting results.

\section*{Appendix}
To prove the existence result of Theorem 2, we start with the following lemma
( \v{S}iljak \cite{Siljak1976-1}, Ladde and \v{S}iljak \cite{Ladde1976}):

\textbf{Lemma 1. \ }If the function $g(t,e)$ satisfies the assumptions
$\left(  A_{1}^{\prime}\right)  ,$ $\left(  A_{3}\right)  $ and $\left(
A_{4}\right)  $, then for any two positive vectors $u>0, v>0,$ $u\neq v,$ there
exist indicies $k,l\in\mathbb{M},k\neq l,$ such that%
\begin{equation}
g_{k}(t,u)<g_{k}(t,v),~\ \ \ g_{l}(t,u)>g_{l}(t,v)
\end{equation}
for each fixed $t\in\mathcal{T}$\ and all $u,v\in\mathbb{R}_{+}^{m}.$

\textbf{Proof.} \ Define $\xi_{k}=\max_{{\large i\in}\mathbb{M}}\{u_{i}%
/v_{i}\},$ $\eta_{l}=\min_{{\large i\in}\mathbb{M}}\{u_{i}/v_{i}\}$ for any
pair of vectors $u,v>0.$ \ With each pair $(u,v),$ we associate the pair
$(u_{\ast},u^{\ast})$ given as $u_{\ast}=\xi_{k}^{-1}u,$ $u^{\ast}=\eta
_{l}^{-1}u,$ so that $u_{\ast}\leq v$ and $u^{\ast}\geq v.$ \ That is,
$u_{\ast i}\leq v_{i},$ $i\neq k,$ $u_{\ast k}=v_{k}$ and, likewise,
$u_{i}^{\ast}\geq v_{i},$ $i\neq l,$ $u_{l}^{\ast}=v_{i}.$ Furthermore, since
$u\neq v$, at least for some $i$ we have $u_{\ast i}<v_{i},$ $u_{i}^{\ast
}>v_{i},i\neq k,i\neq l.$ \ Now, from $\left(  A_{1}^{\prime}\right)  $ and
$\left(  A_{3}\right)  $, we have%
\begin{equation}
g_{k}(t,u)=g_{k}(t,u_{\ast})<g_{k}(t,v),~\ \ \ g_{l}(t,u)=g_{l}(t,u^{\ast
})>g_{l}(t,v)
\end{equation}
for all $t\in\mathcal{T}$ , which proves the lemma.

\textbf{Proof of Theorem 2. \ }Uniqueness of $\mathcal{E}$ means that
for any pair of equilibrium values $e^{\prime},e^{\prime\prime}\in\mathcal{C}%
$, $e^{\prime}\neq e^{\prime\prime},$%
\begin{equation}
g(t,e^{\prime\prime})=g(t,e^{\prime})=0\Rightarrow e^{\prime\prime}=\lambda
e^{\prime}%
\end{equation}
for all $t\in\mathcal{T}$ and some $\lambda>0.$ \ Define $\mu=\min
_{{\large i\in}\mathbb{M}}\{e_{i}^{\prime}/e_{i}^{\prime\prime}\},$ where
$e_{i}^{\prime},e_{i}^{\prime\prime}$ are the $i$-th components of the two
equilibria $e_{i}^{\prime},e_{i}^{\prime\prime},$ and $e^{\prime\prime\prime
}=\mu e^{\prime\prime}$. \ Then, we have $e^{\prime\prime\prime}\leq
e^{\prime},$ that is, $e_{i}^{\prime\prime\prime}\leq e_{i}^{\prime},$ $i\neq
l,$ $e_{l}^{\prime\prime\prime}=e_{l}^{\prime},$ and at least for some $i\neq
l,$ $e_{i}^{\prime\prime\prime}<e_{i}^{\prime}.$ \ Assume that the statement
(20) is false. \ That is, $e^{\prime\prime}\neq\lambda e^{\prime}$ for all
$\lambda>0.$ By $\left(  A_{1}^{\prime}\right)  ,$ $\left(  A_{3}\right)  ,$
$\left(  A_{4}\right)  ,$ and $g(t,e^{\prime\prime})=g(t,e^{\prime})=0,$ we
have $0=g_{l}(t,e^{\prime\prime})=g_{l}(t,e^{\prime\prime\prime}%
)<g_{l}(t,e^{\prime})=0,$ which is absurd. This proves the theorem.

\textbf{Remark 2. }\ If we take any pair of vectors\textbf{ }$e^{e}%
,e\in\mathcal{C}$ such that $e^{e}\in\mathcal{E}$, $e\notin\mathcal{E}$, and
use Theorem 2 and inequalities $(18),$ we conclude that $g_{k}(t,e)<0$ and
$g_{l}(t,e)>0$ for all $t\in\mathcal{T}$ and some indicies $k,l\in\mathbb{M}$.

To establish the existence result of Theorem 3 we can replace the assumptions
$\left(  A_{1}^{\prime}\right)  ,$ $\left(  A_{3}\right)  $ and $\left(
A_{4}\right)  $ by the following weaker assumption:

$\mathbf{Assumption}$ $\mathbf{A}_{5}.$ \ $g(t,e^{e})=0\Leftrightarrow
e^{e}\in\mathcal{E}$, and for any $e\in\mathcal{C}$, $e\notin\mathcal{E}$, and
any $e^{e}\in\mathcal{E}$, there exists a pair of indicies $k,l\in\mathbb{M}$,
$k\neq l,$ such that%
\begin{equation}
e_{k}=\max_{{\large i\in}\mathbb{M}}~\{e_{i}\}\Rightarrow g_{k}%
(t,e)<0,~\ \ \ e_{l}=\min_{{\large i\in}\mathbb{M}}~\{e_{i}\}\Rightarrow
g_{l}(t,e)>0
\end{equation}
for all $t\in\mathcal{T}$ .

\textbf{Remark 3. \ }In view of Remark 2, the assumptions ($A_{1}^{\prime}%
$)$,$ $\left(  A_{3}\right)  ,$ $\left(  A_{4}\right)  $ taken together imply
($A_{5}$) but not vice versa.

Now we can prove a slightly stronger result than Theorem 3 as the following:

\textbf{Theorem 5. }\ If the function $g(t,e)$ satisfies the assumption
$\left(  A_{5}\right)  ,$ then there exists a solution $e(t)=e(t;t_{0},e_{0})$
of \textbf{E }for all $(t_{0},e_{0})\in\mathcal{T\,\times C}$ and
$t\in\mathcal{T}_{0}.$

\textbf{Proof. \ }Consider $e_{0}\notin\mathcal{E}$, and $\alpha=e_{k0}%
=\max_{{\large i\in}\mathbb{M}}\{e_{i0}\},\beta=e_{l0}=\min_{{\large i\in M}%
}\{e_{i0}\},$and $\alpha>\beta>0.$ \ Define%
\begin{equation}%
\begin{tabular}
[c]{l}%
$\mathcal{B}^{\shortmid}=\{e\in\mathbb{R}_{+}^{n}:\beta\leq e_{i}\leq
\alpha,~\forall i\in\mathbb{M\}}$\\
$\mathcal{B}^{\shortmid\shortmid}=\{e\in\mathbb{R}_{+}^{n}:\frac{1}{2}%
\beta\leq e_{i}\leq\alpha+\frac{1}{2}\beta,~\forall i\in\mathbb{M\}}$%
\end{tabular}
\end{equation}
and note $\mathcal{B}^{\shortmid}\subset\mathcal{B}^{\shortmid\shortmid}.$
\ For any $\tau>0,$ we define the time interval $\mathcal{T\,}_{1}%
=[t_{0},t_{0}+\tau]$ and the rectangle $\mathcal{T\,}_{1}\times\mathcal{B}%
^{\shortmid\shortmid}.$ By continuity of $g(t,e)$ we can find a number
$\mu_{1}>0$ such that $\left\vert g_{i}(t,e)\right\vert \leq\mu_{1}$ for all
$(t,e)\in\mathcal{T\,}_{1}\times\mathcal{B}^{\shortmid\shortmid}$ and all
$i\in\mathbb{M}$.  By Peano's existence theorem (e.g., Hale \cite{Hale1969}), there
exists at least one solution $e(t)$ for all $t\in\lbrack t_{0},t_{0}%
+\varepsilon_{1}],$ where $\varepsilon_{1}=\min~\{\tau,\alpha/\mu_{1}\}.$
\ Now, either $e(t_{1})\in\mathcal{E}$ for some $t_{1}\in(t_{0},t_{0}%
+\varepsilon_{1}]$, or $e(t)\notin\mathcal{E}$ for all $t\in\lbrack
t_{0},t_{0}+\varepsilon_{1}].$ In the first case, the solution $e(t)$ exists
for all $t\in\mathcal{T}_{0}.$ When the second case takes place, then we
extend successively the solution $e(t)$ beyond the time interval $(t_{0}%
,t_{0}+\varepsilon_{1}].$

Note that for all $(t_{0},t_{0}+\varepsilon_{1}],$ $e(t)\in\mathcal{B}%
^{\shortmid\shortmid}.$ \ In fact, we can show that the same statement holds
for $\mathcal{B}^{\shortmid}.$ \ Since $e_{0}\notin\mathcal{E}$, from ($A_{5}%
$) we have (21) with $e=e_{0},$ which for $t=t_{0}$ implies%
\begin{equation}
g_{k}(t_{0},e_{0})<0,~\ \ \ g_{l}(t_{0},e_{0})>0.
\end{equation}
By continuity of $g(t,e)$ and $e(t),$ we can find a $\delta_{1}>0$ such that%
\begin{equation}
g_{k}[t,e(t)]<0,~\ \ \ g_{l}[t,e(t)]>0,~\ \ \ \ \forall t\in\lbrack
t_{0},t_{0}+\delta_{1}]
\end{equation}
which by integration yields%
\begin{equation}
e_{k}(t)\leq e_{k0},~\ \ \ e_{l}(t)\geq e_{l0},~\ \ \ \ \forall t\in\lbrack
t_{0},t_{0}+\delta_{1}]
\end{equation}
and, in turn, implies that $e(t)\in\mathcal{B}^{\shortmid}$ for all
$t\in\lbrack t_{0},t_{0}+\delta_{1}].$ \ Since $e(t_{0}+\delta_{1}%
)\in\mathcal{B}^{\shortmid\shortmid}$ and also $e(t_{0}+\delta_{1}%
)\notin\mathcal{E}$, by using again ($A_{5}$)$,$ we get%
\begin{equation}
g_{k}[t_{0}+\delta_{1},e(t_{0}+\delta_{1})]<0,~\ \ \ g_{l}[t_{0}+\delta
_{1},e(t_{0}+\delta_{1})]>0,
\end{equation}
and conclude that there exists a $\delta_{2}>0$ such that%
\begin{equation}
g_{k}[t,e(t)]<0,~\ \ \ g_{l}[t,e(t)]>0,~\ \ \ \ \forall t\in\lbrack
t_{0}+\delta_{1},t_{0}+\delta_{1}+\delta_{2}]
\end{equation}
and%
\begin{equation}
e_{k}(t)\leq e_{k}(t_{0}+\delta_{1}),~\ \ \ e_{l}(t)\geq e_{l}(t_{0}%
+\delta_{1}),~\ \ \ \ \forall t\in\lbrack t_{0}+\delta_{1},t_{0}+\delta
_{1}+\delta_{2}].
\end{equation}
Therefore, (24) and (28) imply that $e(t)\in\mathcal{B}^{\shortmid}$ for all
$t\in\lbrack t_{0}+\delta_{1},t_{0}+\delta_{1}+\delta_{2}].$ By continuing
this process, we conclude in finite number of steps that $e(t)\in
\mathcal{B}^{\shortmid}$ for all $t\in\lbrack t_{0},t_{0}+\varepsilon_{1}].$
\ Since $e(t)$ remains in $\mathcal{B}^{\shortmid}$ for the entire interval
$[t_{0},t_{0}+\varepsilon_{1}],$ by using the above arguments we can show that
the solution $e(t)$ can be extended over the interval $[t_{0}+\varepsilon
_{1},t_{0}+2\varepsilon_{1}]$ and, therefore, over the interval $\mathcal{T\,}%
_{1}=[t_{0},t_{0}+\tau].$ \ Furthermore, $e(t)\in\mathcal{B}^{\shortmid}$ for
all $t\in\mathcal{T\,}_{1}.$

Because the solution $e(t)$ $\in$ $\mathcal{B}^{\shortmid}$ during the entire
time interval $\mathcal{T\,}_{1}=[t_{0},t_{0}+\tau],$ it can be extended over
the interval $\mathcal{T\,}_{2}=[t_{0}+\tau,t_{0}+2\tau]$ by choosing
subintervals of $\mathcal{T\,}_{2}$ determined by $\varepsilon_{2}=\min
\{\tau,\alpha/\mu_{2}\},$ where $\mu_{2}$ is defined by the condition
$\left\vert g_{i}(t,e)\right\vert \leq\mu_{2},$ for all $(t,e)\in
\mathcal{T\,}_{2}\times\mathcal{B}^{\shortmid\shortmid}$ and $i\in\mathbb{M}$.
\ Moreover, we can show as before that $e(t)\in\mathcal{B}^{\shortmid}$ for
all $t\in\mathcal{T\,}_{2}.$ \ Therefore, $e(t)\in\mathcal{B}^{\shortmid}$ for
all $t\in\mathcal{T\,}_{1}\cup\mathcal{T\,}_{2}.$ \ Continuing in this way, we
can find a solution staying inside $\mathcal{B}^{\shortmid}$ for all
$t\in\mathcal{T\,}_{0}.$ \ This proves the theorem.

Now, we can establish our main result concerning the attraction of the
equilibrium ray $\mathcal{E}$ as

\textbf{Theorem 6.} \ If the function $g(t,e)$ satisfies the assumption
$\left(  A_{5}\right),$  then solutions $e(t;t_{0},e_{0})$ of $\mathbf{E}$
have the property ($P_{2}$).

\begin{proof}\cite{Siljak1976-1}
We prove this theorem by using a Liapunov-like function
$V:\mathcal{C}\rightarrow\mathbb{R}_{+},$ defined as%
\begin{equation}
V(e)=d(e,\mathcal{E)}\text{,}%
\end{equation}
and $V(e)\in C^{0}(\mathcal{C)}$. \ Function $V(e)$ is Lipschitzian having the
derivative $D^{+}V(e)_{\mathbf{E}}$ with respect to $\mathbf{E,}$ computed as%
\begin{equation}\label{minV}
D^{+}V(e)_{\mathbf{E}}\leq-\min_{{\large i\in}\mathbb{L}}\{\left\vert
g_{i}(t,e)\right\vert \}\text{, \ \ \ \ }\forall(t,e)\in\mathcal{T\,\times
C-E}\text{,}%
\end{equation}
where $\mathbb{L}$ is a non void subset of $\mathbb{M}$, which is the set of
all $i\in\mathbb{M}$ such that $\left\vert g_{i}(t,e)\right\vert >0.$ \ To
show the last inequality, we note that for each $(t,e)\in\mathcal{T\,\times
C-E}$ there is $\lambda_{0}>0$ such that $V(e)$ can be rewritten as%
\begin{equation}
V(e)=\max_{{\large i\in}\mathbb{M}}\{\left\vert e_{i}-\lambda_{0}\right\vert
\}.
\end{equation}
To see this, we note that the distance between a point $e$ and the ray
$\mathcal{E}$ is equal to the distance between the point $e$ and the foot
$e^{0}\in\mathcal{E}$ of the normal drawn from the point $e$ to the ray
$\mathcal{E}$. \ Furthermore, there exists an index set $\mathbb{L}$ such that
$V(e)=\left\vert e_{i}-\lambda_{0}\right\vert $ for all $i\in\mathbb{L}$.
Now, by assumption $\left(  A_{5}\right)  $ we have either $g_{i}(t,e)>0, $or
$g_{i}(t,e)<0,$and, therefore, $\left\vert g_{i}(t,e)\right\vert
>0, i\in\mathbb{L}$. \ By continuity of $g(t,e)$ and for $\Delta t>0$
sufficiently small, we conclude that the index set $\mathbb{L}$ remains invariant.

We now compute $D^{+}V(e)_{\mathbf{E}}$ as follows:%
\begin{equation}
V[e+\Delta tg(t,e)]-V(e)=\left\vert e_{i}+\Delta tg_{i}(t,e)-\lambda
_{0}\right\vert -\left\vert e_{i}-\lambda_{0}\right\vert \text{,
\ \ \ }\forall i\in\mathbb{L}\text{.}%
\end{equation}
There are two cases to be considered: $x_{i}-\lambda_{0}>0$ and $x_{i}%
-\lambda_{0}<0,$for $i\in\mathbb{L}$. \ When $x_{i}-\lambda_{0}>0$, then
$g_{i}(t,e)<0$, and when $x_{i}-\lambda_{0}<0$, then $g_{i}(t,e)>0.$ \ In
either case, the last equation can be rewritten as%
\begin{equation}
V[e+\Delta tg(t,e)]-V(e)=\Delta t\left\vert g_{i}(t,e)\right\vert \text{,
\ \ \ }\forall i\in\mathbb{L}\text{.}%
\end{equation}
When $x_{i}-\lambda_{0}>0,$ then $g_{i}(t,e)<0,$ and when $x_{i}-\lambda
_{0}<0,$ then $g_{i}(t,e)>0.$ \ Hence, from (33) we get
\begin{equation}\label{VDiff}
V[e+\Delta tg(t,e)]-V(e)=-\Delta t\left\vert g_{i}(t,e)\right\vert ,\text{
\ \ \ }\forall i\in\mathbb{L}\text{,}%
\end{equation}
and finally (\ref{VDiff}) establishes (\ref{minV}).

The second part of the proof consists in proving that the Liapunov-like
function $V(e)=d(e,\mathcal{E)}$ with inequality (\ref{minV}) implies property
\textbf{(}$P_{2}$)$.$ \ Let $e(t)=e(t;t_{0},e_{0})$ be any solution of the
system $\mathbf{E}$ for $(t_{0},e_{0})\in\mathcal{T\,\times C}$. \ Then,
$e(t)\in\mathcal{B}^{\shortmid}$ fot all $t\in\mathcal{T\,}_{0}.$ \ Set%
\begin{equation}
\rho(t)=V[e(t)].
\end{equation}
For sufficiently small $\Delta t>0$, we have%
\begin{equation}%
\begin{array}
[c]{ll}%
\rho(t+\Delta t)-\rho(t) & =V[e(t+\Delta t)]-V[e(t)]\\
& =V[e(t+\Delta t)]-V(e(t)+\Delta tg[t,e(t)])\\
& ~\ \ +V(e(t)+\Delta tg[t,e(t)])-V[e(t)].
\end{array}
\end{equation}
By using the fact that function $V(e)$ is Lipschitzian, we immediately obtain
\begin{equation}\label{Dr}
D^{+}\rho(t)_{\mathbf{E}}\leq\min_{{\large i\in}\mathbb{L}}\{\left\vert
g_{i}(t,e)\right\vert \}\text{, \ \ \ \ }\forall t\in\mathcal{T\,}_{0}.
\end{equation}

We now proceed to establish property ($P_{2}$) by contradiction. \ For some
$\varepsilon>0$, $(t_{0},e_{0})\in\mathcal{T\,\times C}$, there exists
$t_{1}>t_{0}$ and a sequence $\{t_{k}\},$ $t_{k}>t_{1},$ $t_{k}\rightarrow
+\infty,$ $k\rightarrow+\infty$, such that $d[e(t_{k}),\mathcal{E]}%
=\varepsilon$ and $d[e(t),\mathcal{E]>}~\varepsilon$ for $t\in(t_{k}%
,t_{k+1}).$ \ Let us introduce
\begin{equation}
\mathcal{B}^{\shortmid\shortmid\shortmid}=\{e\in\mathcal{B}^{\shortmid
}:d(e,\mathcal{E]\geq}~\ \varepsilon\mathbb{\}}\text{,}%
\end{equation}
which is a compact set. \ For any $t\in\mathcal{T\,}$, and any fixed
$\tilde{e}\in\mathcal{B}^{\shortmid\shortmid\shortmid}$, there is an index
subset $\mathbb{L\subset M}$, such that the function%
\begin{equation}
\theta(t,\tilde{e})=\min_{{\large i\in}\mathbb{L}}\{\left\vert g_{i}%
(t,\tilde{e})\right\vert \}>0.
\end{equation}
By continuity of $\theta(t,\tilde{e}),$ there exists a neighborhood
$\mathcal{N(}\tilde{e})$ of $\tilde{e}\in\mathcal{B}^{\shortmid\shortmid
\shortmid}$ such that $\theta(t,e)>0,$ for all $e\in\mathcal{N(}\tilde{e}).$
\ Let $\mathcal{U=\{N(}\tilde{e}):\tilde{e}\in\mathcal{B}^{\shortmid
\shortmid\shortmid}\}$ be an open cover of $\mathcal{B}^{\shortmid
\shortmid\shortmid}$. \ Since $\mathcal{B}^{\shortmid\shortmid\shortmid}$ is
compact, by Heine-Borel Theorem (Royden \cite{Royden1968}), we can extract a finite
subcover \{$\mathcal{N}$ ($\tilde{e}_{1}),\mathcal{N}$ ($\tilde{e}_{2}%
),$\ldots, $\mathcal{N}$($\tilde{e}_{m})\},$ where to each $\mathcal{N}%
$($\tilde{e}_{j})$ there corresponds an index subset $\mathbb{L}_{j}$ and
function%
\begin{equation}
\theta_{j}(t,e)=\min_{{\large i\in}\mathbb{L}_{{\large j}}}\{\left\vert
g_{i}(t,e)\right\vert \}.
\end{equation}
We define $\psi(t,e)=\min_{j}\{\theta_{1}(t,e),\theta_{2}(t,e),$\ldots,
$\theta_{m}(t,e)\},$ and note that $\psi(t,e)\in C^{(0,0)}(\mathcal{T\,}%
\times\mathcal{B}^{\shortmid\shortmid\shortmid}),$ and $\psi(t,e)>0,$ for all
$(t,e)\in\mathcal{T\,}\times\mathcal{B}^{\shortmid\shortmid\shortmid}.$
Therefore, we can take $\inf_{e\in\mathcal{B}^{\shortmid\shortmid\shortmid%
%TCIMACRO{\U{b4}}%
%BeginExpansion
\acute{}%
%EndExpansion%
%TCIMACRO{\U{b4}}%
%BeginExpansion
\acute{}%
%EndExpansion%
%TCIMACRO{\U{b4}}%
%BeginExpansion
\acute{}%
%EndExpansion
}}\psi(t,e)=\varphi(t),$ and $\varphi(t)\in C^{0}(\mathcal{T\,)}$ since
$\mathcal{B}^{\shortmid\shortmid\shortmid}$ is compact.

Now, from inequality (\ref{Dr}), we can obtain%
\begin{equation}
D^{+}\rho(t)_{\mathbf{E}}\leq-\varphi(t),
\end{equation}
where $e(t)\in\mathcal{B}^{\shortmid\shortmid\shortmid},t\in\lbrack
t_{k},t_{k+1}].$ \ Integrating this inequality from $t_{k}$ to $t_{k+1},$ and
using the definitions of $V(e)$ and $\rho(t),$ we get%
\begin{equation}
0=\rho(t_{k+1})-\rho(t_{k})\leq-%
%TCIMACRO{\dint \limits_{t_{k}}^{t_{k+1}}}%
%BeginExpansion
{\displaystyle\int\limits_{t_{k}}^{t_{k+1}}}
%EndExpansion
\varphi(\tau)d\tau<0,
\end{equation}
which is absurd. Therefore, the proof of the theorem is complete.
\end{proof}

\section*{Acknowledgment}
My first thanks go to my longtime colleague, co-author and friend Professor Gangaram S. Ladde, Department of Mathematics and Statistics, University of South Florida, for his generous help in writing the first phase of this paper.  I am also exceedingly grateful to my colleague Professor Mohammad A. Ayoubi, Mechanical Engineering Department, Santa Clara University, for his help in completing the paper.

\end{document}